\documentstyle[twocolumn, eqsecnum,aps,graphicx]{revtex}
\begin{document}

\title{\bf Butterfly hysteresis loop and dissipative spin reversal in the
S=1/2, V$_{15}$ molecular complex}

\author{I. Chiorescu$^a$, W. Wernsdorfer$^a$, A. M\"uller$^b$, H.
B\"ogge$^b$ and B. Barbara $^a$.}

\address{$^a$ Laboratoire de Magn\'etisme Louis N\'eel, CNRS, BP 166,
38042-Grenoble, France.\\
 $^b$ Fak\"ultat f\"ur Chemie, Universit\"at Bielefeld, D-33501 Bielefeld,
Germany. }

\date{\today}

\maketitle

\begin{abstract}
Time resolved magnetization measurements have been performed on a spin 1/2
molecular complex, so called V$_{15}$. Despite the absence of a barrier,
magnetic hysteresis is observed over a timescale of several seconds. A
detailed analysis in terms of a dissipative two level model is given, in
which fluctuations and splittings are of same energy. Spin-phonon
coupling leads to long relaxation times and to a particular "butterfly"
hysteresis loop.

\end{abstract}
\narrowtext

In this letter we study the dynamics of the magnetization reversal of a
molecular crystal made of nanometric molecules with non-interacting $S=1/2$
spins. Despite the absence of energy barrier against spin reversal, this system
shows hysteresis. This result are interpreted in details assuming spin
rotation in a phonon bath, which is different from the situation of large spin
molecules where only the spin bath is believed to be relevant
\cite{NatFriedJMMM,BB,GargStamp,Stamp0}. Resonant phonon transitions are
irrelevant, unless between states at different energies
\cite{BoutronVillainLoss} or in the presence of a transverse field large
enough to create a tunnel splitting of the order of the temperature energy
scale \cite{Belessa}. 

The molecular complex
K$_6$[V$_{15}^{IV}$As$_6$O$_{42}$(H$_2$O)]$\cdot$8H$_2$O (so-called V$_{15}$)
\cite{Muller} is made of molecules with fifteen V$^{IV}$ ions of spin
$S=1/2$, placed in a quasi-spherical layered  structure formed of a triangle,
sandwiched by two hexagons. The symmetry is trigonal (space  group $R\bar3c$,
$a=14.029$~\AA, $\alpha=79.26^{\circ}$, $V=2632$~\AA$^ 3$).  The unit-cell
contains two V$_{15}$ clusters and it is large enough that dipolar
interactions between different  molecules are negligible (a few mK). All
intra-molecular exchange interactions being antiferromagnetic, the total spin
of this molecule is $S=1/2$. Such a small spin has zero energy barrier
and relatively large splitting in zero applied field ($\sim10^{-2}$~K).
 Although spin entanglement results in 2$^{15}$ eigenstates per molecule, the
magnetization curves will be interpreted in terms of a dissipative two level
model \cite{LeggettGrifoni,Zener,MiyaKayaStern}. 

\begin{figure}
\begin{center}\includegraphics[width=8.1cm]{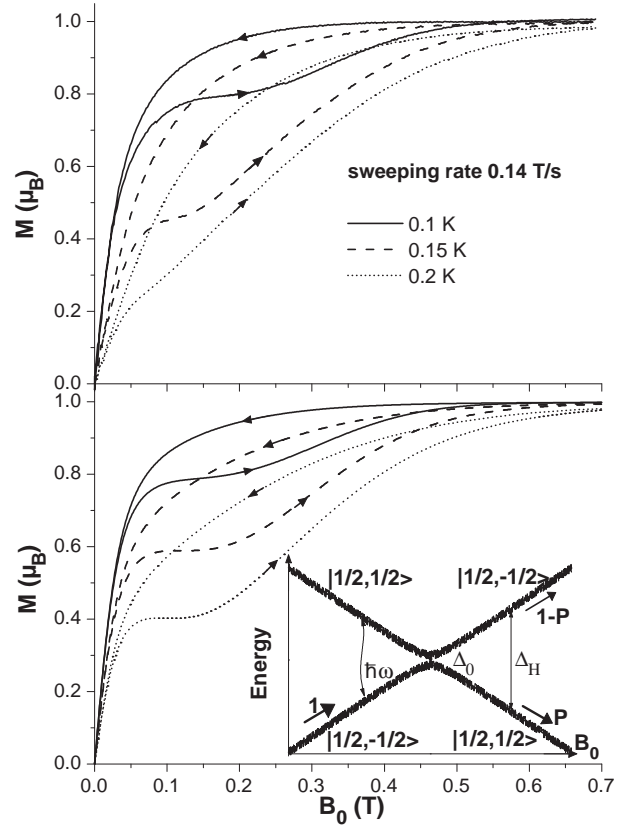}\end{center}
\caption{Measured (\textbf{a}$-$\emph{top}) and calculated
(\textbf{b}$-$\emph{bottom}) hysteresis loops for three temperatures and for
a given field sweeping rate 0.14~T/s. The plateau is more pronounced at low
T.  The inset is  a schematic representation of a two-level system $S_Z=\pm1/2$
with repulsion due to non-diagonal matrix elements. In a swept field the
switching probability $P$ is given by the Landau-Zener formulae (see text).
The two levels are broadened by the hyperfine fields and the absorption
or the emission of phonons can switch the polarization state of spins.}
\label{3Temps} \end{figure}

\begin{figure}
\begin{center}\includegraphics[width=8.1cm]{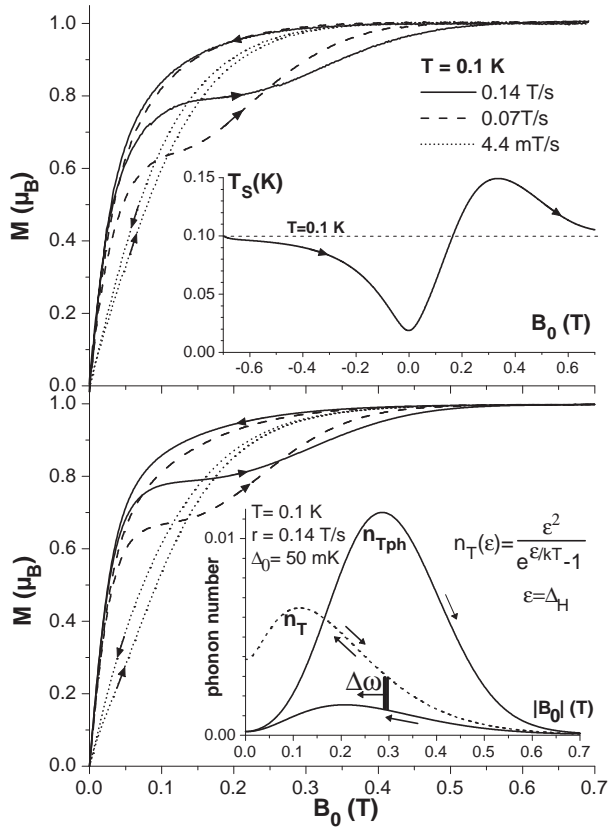}\end{center}
\caption{Measured (\textbf{a}$-$\emph{top}) and calculated
(\textbf{b}$-$\emph{bottom}) hysteresis loops for three field sweeping rates
at $T=0.1$~K. The observed plateau is more pronounced at high sweeping rate.
The equilibrium curve can be approximated by the median
of the two branches of the low sweeping rate hysteresis loop  (dotted curve). 
In the \emph{top} inset is plotted the spin and phonon temperature $T_S=
T_{ph}$ for $T=0.1$~K and $r=0.14$~T/s, when the field is swept from negative
values. $T_S$ decreases until zero-field and then increases linearly within the
plateau region. Then it overpasses the bath temperature to finally reach the
equilibrium. In the \emph{bottom} inset the calculated number of phonons with
$\hbar\omega=\Delta_H$ is plotted vs. the sweeping field modulus (note the
arrows) at equilibrium ($T_{ph}=T_S=T$, dashed line) and
out-of-equilibrium ($n_{T_{ph}}=n_{T=T_S}$, $r=0.14$~T/s, black line). The
difference between the two curves (thick segment $\Delta\omega$) suggests the
moving hole in the phonon distribution, while their intersection gives the
plateau intercept of the equilibrium magnetization curve.}

\label{3rates} \end{figure}

Time-resolved magnetization measurements were performed with the micro-SQUID
technique ($50-400$~mK, $0-0.7$~T/s) \cite{PRLFeBa}. In order to maximaze
thermal contact with the bath, we choose a sample holder made by greece and
silver powder and a small crystal of the V$_{15}$ ($\sim50$~$\mu$m). As an
example we give a few hysteresis loops in Fig.~\ref{3Temps}a and
Fig.~\ref{3rates}a (only the positive parts are represented, the other ones
being rigorously symmetrical). When the field increases, coming from the
negative saturation, the magnetization curve passes through the origin of the
coordinates, reaches a plateau and then approaches saturation. This leads to a
winged hysteresis loop characterized by the absence of irreversibility near
zero field. Nevertheless, the initial susceptibilities being larger the faster
sweeping field, the magnetization is out of equilibrium also near zero field
where it appears to be reversible. 

The wings depend sensitively on temperature $T$ and field sweeping rate $r$.
In Fig.~\ref{3Temps}a, where three hysteresis loops are presented at three
different temperatures for a given sweeping rate, the plateau is higher and
more pronounced at low temperature. The same tendency is observed at a given
temperature and faster sweeping rates (Fig.~\ref{3rates}a). When compared to
its equilibrium limit (dotted curve in Fig.~\ref{3rates}), each magnetization
curve shows a striking feature: the plateau intersects the equilibrium curve
and the magnetization becomes smaller than at equilibrium. Equilibrium is then
reached in higher fields near saturation.

In order to interpret this magnetic behavior of the  V$_{15}$ molecules, we
will analyse how the level occupation numbers vary in this two level system
(see Fig.~\ref{3Temps}b inset) when sweeping an external field. In the
absence of dissipation, a 2-level model is well described by the bare
Landau-Zener model, in the adiabatic or non-adiabatic case (low or high
sweeping rates). The probability for the
$|1/2,-1/2\rangle\leftrightarrow|1/2,1/2\rangle$ transition is
$P=1-\exp(-\pi\Delta_0^2/4\hbar\mu_Br)$. In such a Landau-Zener  transition,
the plateaus of Fig.~\ref{3rates} should decrease if the sweeping rate
increases, which is contrary to the experiments. Taking the typical value
$r=0.1$~T/s and the zero-field splitting $\Delta_0\cong 0.05$~K
\cite{hyp,Dobro,ICM,Carter}, one gets a ground state switching probability
very close to unity: in the absence of dissipation the spin 1/2 must
adiabatically follow the field changes. Extremely large sweeping rates ($\approx
10^9$~T/s) would be needed to get into the quantum non-adiabatic regime $P<1$.
The mark of the V$_{15}$ system is that the dissipative spin-phonon coupling
is acting also near zero applied field because $\hbar\omega\approx\Delta_0$ is
of the order of the bath temperature, which is not the case for large spin
molecules where $\Delta_0<<k_BT$. The spin temperature $T_S$ is such that 
$n_1'/n_2= \exp(\Delta_H/k_BT_S)$, where
$\Delta_H=\sqrt{\Delta_0^2+(2\mu_BB_0)^2}$ is the two levels field-dependent
separation, and $n_{1,2}$($n_{1,2eq}$) the out of equilibrium (equilibrium)
level occupation numbers. In the  magnetization curves at 0.1~K
(Fig.~\ref{3Temps},\ref{3rates}a), the spin temperature is significantly lower
than the bath temperature $T$ ($n_{1}>n_{1eq}$, $T_S<T$) between $-0.3$~T
(when the magnetization curve departs from the equilibrium one) and 0.15~T
(the field at which the magnetization curve intersects the equilibrium one).
After this intersept $T_S$ is larger than the bath temperature
($n_{1}<n_{1eq}$, $T_S>T$), and at sufficiently high fields (about 0.5~T) it
reaches the equilibrium value ($n_{1}=n_{1eq}$, $T_S=T$).

	In a direct process, the spins at the temperature $T_S$ should relax to
the phonons temperature within a timescale $\tau_1$, the phonons being at the
bath temperature. However, even with a silver sample holder, it is not
possible to maintain the phonon temperature equal to the temperature of the
bath. This is because in V$_{15}$ below 0.5 K, the heat capacity of the
phonons $C_{ph}$ is very much smaller than that of the spins $C_S$, so that
the energy exchanged between spins and phonons will very rapidly adjust the
phonons temperature $T_{ph}$ to the spin one $T_S$. Furtheremore, the energy
is transfered from the spins only to those phonon modes with
$\hbar\omega=\Delta_H$ (within the resonance line width). The number of such
lattice modes being much smaller than the number of spins, energy transfer
between the phonons and the sample holder must be very difficult, a phenomenon
known as the phonon bottleneck \cite{VanVleckStev}. Following \cite{A&B}, the
number of phonons per molecule available for such resonant transitions is
$n_{T}=\int_{\Delta\omega}\sigma (\omega) d\omega/(\exp(\hbar\omega/kT)-1)$,
where $\sigma (\omega)d\omega=3V\omega^2d\omega/(2\pi^2v^3)=$ number of phonon
modes between $\omega$ and $\omega+d\omega$ per  molecule of volume $V$, $v$
is the phonon velocity and $\Delta\omega$ is the transition linewidth due to
fast hyperfine field fluctuations (they broden both energy
levels)\cite{Stamp}. Taking the typical values $v\approx 3000$~m/s,
$T\approx10^{-1}$~K and $\Delta\omega\approx 5\cdot10^2$~MHz we find $n_{T}$ 
of the order of $\approx10^{-6}$ to $10^{-8}$ phonons/molecule. Such a small
number of phonons is very rapidly absorbed, burning a hole of width
$\Delta\omega$ in the phonon density of states at the energy
$\hbar\omega=\Delta_H$ \cite{VanVleckStev}. If this phonon density of states
does not equilibrate fast enough, the hole must persist and move with the
sweeping field, leading to a phonon bottleneck.

Now this description will be made quantitative. For a given splitting
$\Delta_H$, the time evolution of the two levels populations $n_{1,2}$ and of
the phonon numbers $n_{T_{ph}}$ at $T_{ph}$ obeys the set of two differential
equations \cite{A&B}: (i) $-\dot n_1=\dot n_2=P_{12}n_1-P_{21}n_2$ and (ii)
$\dot n_{T_{ph}}=-(n_{T_{ph}}-n_T)/\tau_{ph}-P_{12}n_1+P_{21}n_2$, where
$P_{12,21}$ are the transition probabilities between the two levels (they are
themselves linear functions of $n_{T_{ph}}$) and $\tau_{ph}\approx L/2v$ is the
phonon-bath relaxation time ($L$ is the sample size). Using the notations
$x=(n_1-n_2)/(n_{1eq}-n_{2eq}), y=(n_{T_{ph}}-n_T)/(n_T+n/2)$ with
$n=\int_{\Delta\omega}\sigma (\omega) d\omega$ we get: (i) $\dot
x=(1-x-xy)/\tau_1$ and (ii) $\dot y=-y/\tau_{ph}+b\dot x$, where
$b=C_S/C_{ph}$ and $1/\tau_1=P_{12}+P_{21}$ the direct spin-phonon relaxation
time. By solving numerically this system for typical values, e.g.
$\tau_1=10^{-2}$~s, $\tau_{ph}<10^{-6}$~s,  $b>10^5$, we can see that
$T_{ph}\rightarrow T_S\not=T$ (phonon bottleneck) very rapidly, as
expected.This leads to  $y=1/x-1$ and the second equation of the differential
system becomes $\dot x=(x-x^2)/(1+bx^2)/\tau_{ph}$. In the limit $b>>1$ (in
our case $b\approx10^8-10^{10}$) this equation has the solution: 
\begin{equation} -t/b\tau_{ph}=x-x_0+\ln((x-1)/(x_0-1)), \label{trelax}
\end{equation}
where $x_0=x(t=0)$ and $b\tau_{ph}$ is the spin-phonon recovery relaxation
time ($T_{ph}=T_S\rightarrow T$). When the system is not far from
equilibrium ($x\sim1$), we get an exponential decay of the magnetization, 
with the same time constant $\tau_H=b\tau_{ph}$.  For a spin 1/2 system
\cite{A&B}:  \begin{equation} 
\tau_H=\alpha\frac{\tanh^2(\Delta_H/2k_BT)}{\Delta_H^2}, 
\label{tauform}
\end{equation}
with $\alpha=2\pi^2\hbar^2v^3N\tau_{ph}/3\Delta\omega$ ($N$ the molecule
density).

 The dynamical magnetization curves calculated in this model are given
Fig.~\ref{3Temps}b and Fig.~\ref{3rates}b. We started from equilibrium
($x_0=1$) in large negative fields. Then we let the system relax for a very
short time $\delta t$ and we calculated $x(\delta t)$ using Eq.~\ref{trelax}.
This value was taken as the initial value for the next field (the field step
is $r\delta t$). The parameters have been chosen to mimic the measured curves
of Fig.~\ref{3Temps}a and Fig.~\ref{3rates}a \cite{AD}. The obtained
similarity supports the possibility of the phonon bottleneck effect at the
timescale of a few 0.1 s. In the Fig.~\ref{3rates}a inset, we show the
variation of the calculated spin-phonon temperature $T_S$ for $T=0.1$~K and
$r=0.14$~T/s. We can note a linear variation in the plateau region (small
positive fields, $n_1/n_2\approx cst.$), after a cooling in negative fields.
The slope of this quasi-adiabatic linear region varies with the bath
temperature and sweeping rate and gives the plateau dependence on these two
parameters (see Figs.~\ref{3Temps},~\ref{3rates}). 

\begin{figure}
\begin{center}\includegraphics[width=8.1cm]{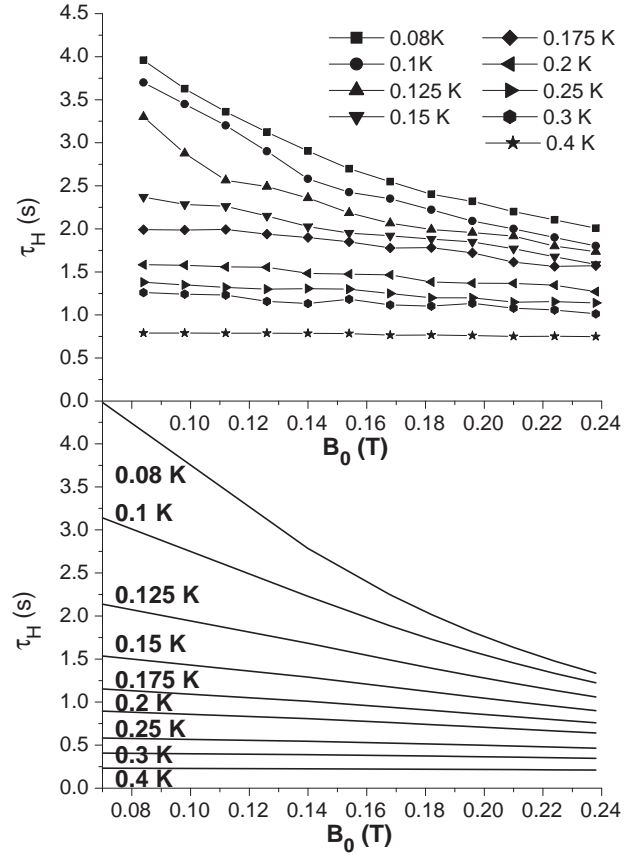}\end{center}
\caption{The relaxation times $\tau_H$, measured
(\textbf{a}$-$\emph{top}) and calculated (\textbf{b}$-$\emph{bottom}, same
parameters as in Fig.~\ref{3Temps}, \ref{3rates}b).}
 \label{taufit}
\end{figure}

In the Fig.~\ref{3rates}b inset we show the calculated field evolution of the
number of phonons at energy $\hbar\omega=\Delta_H$ at equilibrium
($T_{ph}=T_S=T$, dashed line) and out-of-equilibrium ($n_{T_{ph}}=n_{T=T_S}$,
$r=0.14$~T/s, black line). The difference between the two curves (thick segment
$\Delta\omega$) suggests the moving hole in the phonon distribution, while
their intersection gives the plateau intercept of the equilibrium
magnetization curve (above which the hole dissapears and $T_{ph}=T_S>T$). Let
note that in zero field the system is out-of-equilibrium even if magnetization
passes through the origin of coordinates (without a barrier, the switch
between $+1/2$ and $-1/2$ follows the level structure shown Fig.~\ref{3Temps}
inset ). At larger fields, in the plateau region, $n_1/n_2\approx cst.$ at
timescales shorter than $\tau_H=b\tau_{ph}$ (Eq.~\ref{tauform}), even after
the plateau crosses the equilibrium curve. Equilibrium is reached when
$\tau_H$ becomes small enough.

Furthermore, we measured the relaxation of the magnetization of our crystal
at different fields and temperatures, along the plateau region. The
relaxation curves compared well to exponential decay and the obtained
relaxation times are presented in Fig.~\ref{taufit}a. The comparison with
those calculated (Fig.~\ref{taufit}b) is acceptable. But we noted that a
direct fit to Eq.~\ref{trelax} would necessitate larger values for $\alpha$ and
$\Delta_0$ ($\approx0.4-0.6$~sK$^2$ and $\approx0.2-0.3$~K). Note that in
V$_{15}$ we have $b\tau_{ph} > \tau_1$ and this leads to the phonon bottle-neck
regime. However, in other systems one might have $b\tau_{ph} < \tau_1$ in which
case the phonons would be at equilibrium but still with a butterfly
hysteresis loop ($\tau_H$ is a linear combination of $\tau_1$ and $b\tau_{ph}$
\cite{A&B}). This type of hysteresis loop is general and characterizes
dissipative spin reversal in the absence of barrier.

In conclusion, the V$_{15}$ molecular complex constitutes an example of
dissipative two-levels system \cite{LeggettGrifoni} of mesoscopic size. The
total spin 1/2 being formed of a large number of interacting spins, its
splitting results from the structure itself of the molecule (intra-molecular
hyperfine and Dzyaloshinsky-Moriya couplings) and it is rather large (a
fraction of Kelvin)\cite{hyp}. In V$_{15}$ and in other low-spin systems,
splittings must be much larger than in large-spin molecules where the presence
of energy barriers lowers them by orders of magnitude (e.g. $10^{-11}$~K in
Mn$_{12}$ \cite{NatFriedJMMM,BB}).This is the reason why spin-phonon
transitions within the tuneling gap are important in low-spin molecules and not
relevant in high-spin ones, unless a large transverse field is applied 
\cite{Belessa} (it increases the tunnel splitting and probability) in which
case we would also expect similar phenomena.

\acknowledgements
We are very pleased to thank P.C.E. Stamp, S. Myashita, I. Tupitsyn, A. K.
Zvezdin, V. Dobrovitski, H. De Raedt and A. J. Leggett for useful discussions
and E. Krickemeyer, P. K\"ogerler, D. Mailly, C. Paulsen and J. Voiron for
on-going collaborations.

\noindent \emph{corresponding author: ichiores@polycnrs-gre.fr}


\begin{references}
\bibitem{hyp} For an isolated Kramers spin 1/2, $\Delta_0=0$. However, the
total spin 1/2 of V$_{15}$ comes from 15 coupled spins and different
intra-molecular couplings such as hyperfine ($A\vec{I}\cdot\vec{S}, I=7/2,
A\approx10$~mK \cite{Carter}) and Dzyaloshinsky-Moriya interactions
\cite{Dobro,ICM} could generate a splitting. In particular, the two $S=1/2$
low-lying degenerate doublets and the $S=3/2$ first excited quartet could be
slightly mixed by D-M interactions, removing the degeneracy of the $S=1/2$
doublet. A value $\Delta_0\approx50$~mK is strongly supported by the
experiment-to-model comparison in Figs.~\ref{3Temps} and \ref{3rates}.
\bibitem{Stamp} The fast hyperfine fluctuations are characterized by the
transvers nuclear relaxation time $T_2$ associated with the dipolar
internuclear interactions and the total spread of the
energy is given by hyperfine couplings \cite{Stamp0}.
\bibitem{AD} $\alpha=0.15$~sK$^2$ ($2.9\cdot10^{-47}$~sJ$^2$),
$\Delta_0=50$~mK within a precision range of $\approx20\%$. Taking
$L\sim30-50~\mu m, N\sim 10^{27}$~m$^{-3},
\Delta\omega\sim5\cdot10^8$~s$^{-1}$, one gets a phonon velocity $v\approx
2800-3600$~m/s which is quite a reasonable value.
\end{references}
\end{document}